# Introducing OpenMP Tasks into the HYDRO Benchmark

Jérémie Gaidamour[a], Dimitri Lecas[a], Pierre-François Lavallée[a]

[a]*IDRIS/CNRS, Campus universitaire d'Orsay, rue John Von Neumann, Bâtiment 506, F-91403 Orsay, France*

**Abstract**

The HYDRO mini-application has been successfully used as a research vehicle in previous PRACE projects [6]. In this paper, we evaluate the benefits of the tasking model introduced in recent OpenMP standards [9]. We have developed a new version of HYDRO using the concept of OpenMP tasks and this implementation is compared to already existing and optimized OpenMP versions of HYDRO.

## 1. Introduction

HYDRO [1] is a mini-application which implements a simplified version of RAMSES [2], a code developed to study large scale structure and galaxy formation. HYDRO uses a fixed rectangular two-dimensional space domain and solves the compressible Euler equations of hydrodynamics using a finite volume discretization of the space domain and a second-order Godunov scheme with splitting direction technique [3]. Fluxes at the interface of two neighbouring computational cells are computed using a Riemann solver [4]. This numerical scheme is fully described in [5].

The solver uses a second order scheme and therefore, updating the grid values at each time step requires the knowledge of its eight neighbour cells. The characteristic of the algorithm is that knowledge of the neighbours in both directions of the grid is not required at the same time. At each time step, a one-dimensional Godunov scheme is performed successively along each direction of the space domain. When processing a direction, the computation of new grid cell values only depends on the values of the neighbour cells along this direction as described in Figures 1 and 2. There is no dependency in the computation of two neighbouring cells who share an interface along the other direction.

## 2. OpenMP "Fine-Grain" version

Depending on the direction of the Godunov scheme, the iteration updating of either each column or each row of the two-dimensional space domain can be performed in parallel. This key property is used in the OpenMP "Fine-Grain" version of HYDRO as described in [6]: The loop through the rows (or columns in the other direction) is parallelized using an OpenMP DO construct in the Godunov routine (see Figures 3 and 4).

Unfortunately, the scalability of this version is limited as core counts increase. One reason for this is the lack of data locality: During the two successive 1D sweeps of the same time step, one thread is responsible for the computation of different chunks of the global domain.

## 3. OpenMP "Coarse-Grain" version

The OpenMP "Coarse-Grain" version of HYDRO described in [6] exploits data locality by using a 2D domain decomposition of the global domain. The implementation is very similar to the MPI version of HYDRO as each thread is responsible for the computation of a local subdomain. The main difference is that the OpenMP version

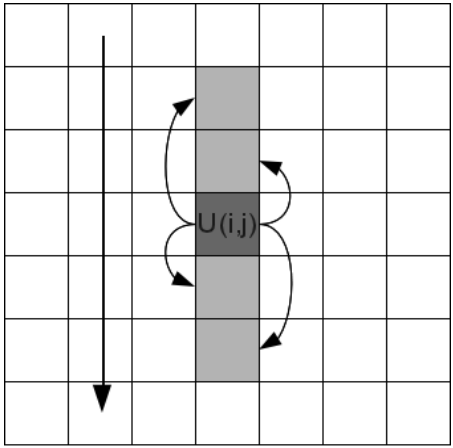

$U_n(i,j) = function( U_{n-1}(i-2,j), U_{n-1}(i-1,j),$
$U_{n-1}(i,j), U_{n-1}(i+1,j), U_{n-1}(i+2,j) )$

**Figure 1:** Column-wise update of U(i,j).

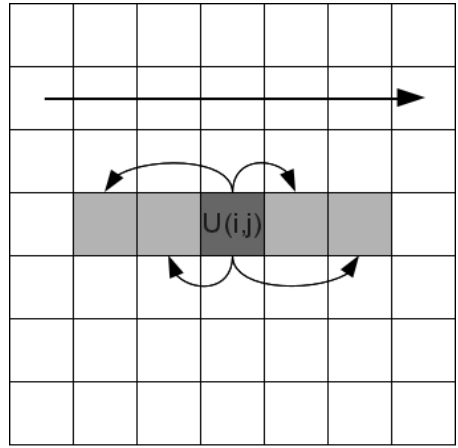

$U_n(i,j) = function( U_{n-1}(i,j-2), U_{n-1}(i,j-1),$
$U_{n-1}(i,j), U_{n-1}(i,j+1), U_{n-1}(i,j+2) )$

**Figure 2**: Row-wise update of U(i,j).

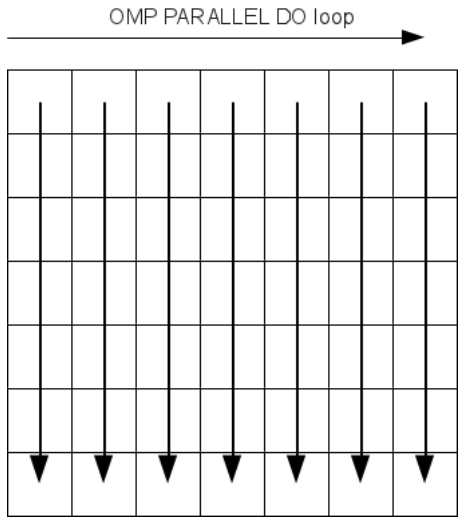

**Figure 3:** 1D Godunov time step routine in the column direction. Arrows represent parallel processing.

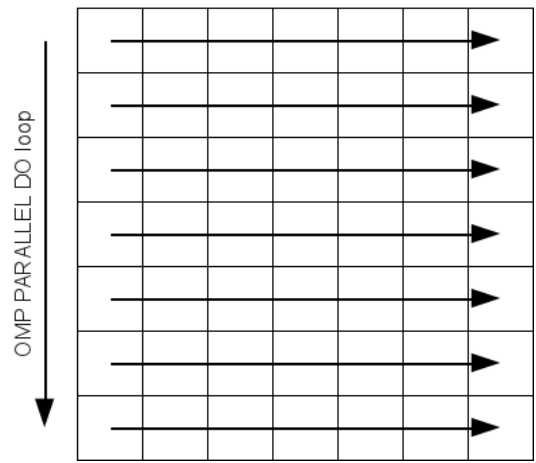

**Figure 4:** 1D Godunov time step routine in the row direction.

does not require extra storage for ghost cells between subdomains because neighbour domains can be accessed directly using the shared memory. The algorithm used on each subdomain is described in Figure 5. The subdomain is processed one column (or row) at a time and the update process takes place using temporary buffers corresponding to the processed subdomain column (or row).

For the interface cells, synchronization between threads is needed, as a thread must read (step (1) of the algorithm in Figure 5) the initial values of its neighbour subdomains (ghost cells) before neighbour threads update them (step (3) of the algorithm). These interactions between threads are shown in Figures 6 and 7. In the MPI version, this synchronization is done implicitly by communicating the ghost cells between MPI processes before starting the actual computation. In the OpenMP version, this synchronization between threads must be done explicitly.

Before step (3), threads have to wait until step (1) of the two neighbour domains is complete. This thread-level synchronization is a barrier involving threads grouped in threes: For each column, a thread is blocked just before the writing step until its two neighbours read the initial values they need for their own computation. The implementation of this synchronization is described in [6]: Threads share a buffer of integer values. There is one integer by thread and values identify the last column read by each thread. Each thread updates its

> U(i , j) is the 2D grid buffer for conservative variables.
>
> **For each** time step $n$ :
> - Apply boundary conditions
> - **PARALLEL LOOP**: For each column $j$ of a subdomain:
>     (1) **READ** - Copy the values of the $j$ column into a 1D temporary buffer.
>         The buffer holds conservative variable values of the previous time step ($U_{n-1}(:,j)$).
>     (2) **COMPUTE** - Compute the new grid values ($U_n(:,j)$) only from the temporary buffer
>         (ie: compute primitive variables, solve Riemann problem at cell interfaces and compute incoming fluxes).
>     (3) **WRITE** - Copy $U_n(:,j)$ in global U.
> - **PARALLEL LOOP**: For each row $i$ of a subdomain:
>     […]

**Figure 5:** 1D Godunov time step routine in the column and row direction (pseudo-code).

corresponding value after step (1). The barrier before step (3) consists of checking the buffer values of the two neighbour domains. The implementation involves flushing the buffer with the OpenMP FLUSH directive (after every value updates and also inside the busy-waiting loop at the barrier).

The synchronization compels a thread to be, at most, one column ahead of its neighbours for the computational part, but has to wait for the other threads to start processing the same column before being allowed to move to the next loop iteration. This synchronization between threads might seem excessive as it forces them to work in concert at the column level but in practice, there is not any waiting at the barrier because reading steps are overlapped with computations. The workload is well-balanced and a thread waits at the barrier only if it is able to complete its own reading and computation step of the column before its two neighbours complete their reading step.

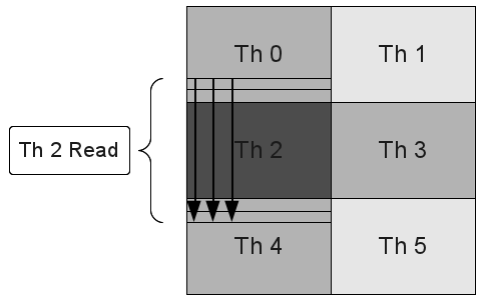 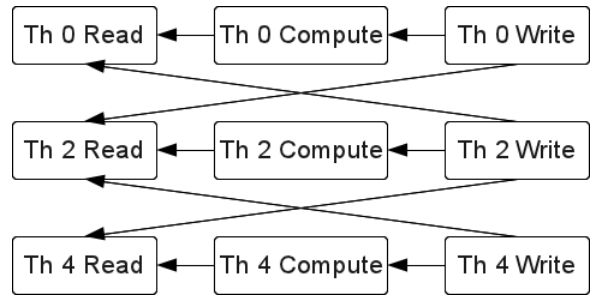

**Figure 6:** Godunov update (column wise) performed by the second thread.

**Figure 7**: Dependency graph between threads in the OpenMP "Coarse-Grain" version.

## 4. OmpSs version of HYDRO

In reference [7], the authors present a first attempt at introducing task-based parallelism into HYDRO. The study uses the OmpSs programming model [8], an extension of the OpenMP standard developed at the Barcelona Supercomputer Center to provide support for asynchronous parallelism and heterogeneous architectures. It allows expressing an algorithm as a set of tasks to be executed by available threads; dependencies between tasks can be expressed to enforce an evaluation order of the tasks. In this paper, the authors use the OpenMP "Fine-Grain" version of HYDRO as a starting point. As OmpSs does not provide a parallel DO loop construct, iterations of the loops are divided into chunks, each one containing Hstep iterations, and an independent task is created for each chunk. This simply limits the amount of tasks that are generated and indeed implements an equivalent of an OMP DO SCHEDULE(..., Hstep) instruction where chunks are scheduled at runtime by the OmpSs task scheduler. Each task is responsible for the computation of a column block (or a row block) and the computation that is done one column (or row) at a time for each block can be reorganized. Authors of the OmpSs version attempted to extract parallelism from each column (or row) block by interchanging nested loops, splitting the computation into blocks on the second direction and generating one task for each subroutine involved in the algorithm. But as the subroutines still form a linear chain of dependencies, they obtained no real benefit [7].

## 5. Introducing OpenMP tasks in HYDRO

The version 4.0 of the OpenMP specifications [9] introduces the concept of tasks with dependencies and the new OpenMP directives are very similar to those provided by the OmpSs programming model. The goal of this paper is to introduce OpenMP tasks in the "Coarse-Grain" version of HYDRO. We aim at implementing the subdomain synchronization we described previously using task dependencies.

In the original version of the code, there is one thread per subdomain and each thread executes the same workflow on different pieces of the domains (achieving data parallelism). Replacing threads with tasks required some code refactoring. The code was reorganized into computational kernels corresponding to the different steps of a subdomain computation: In particular, the Godunov routine was split into three kernels (ie: READ/COMPUTE/WRITE) so that the dependencies between these kernels could be described with OpenMP directives. Private variables corresponding to the subdomain information (ie: domain coordinates and neighbours) and temporary buffers shared by successive kernels were moved to the shared memory. We obtained a program where threads and subdomains are untied and where kernels apply to the subdomain passed as input argument. Incremental reorganization of data and routine interfaces was possible without deep knowledge of the algorithmic details.

Henceforth, we focus on the column direction but the discussion also applies to the Godunov update in the row direction. In the original version of the column-wise Godunov sweep, each thread is working on one subdomain column at a time. If we use one (or more) task per subdomain column, the high number of tasks might incur an important CPU overhead due to the task management. The number of tasks can be reduced by expressing the task dependencies for a set of columns. Note that instead of introducing column blocking for each subdomain, this sectioning can be performed by refining the domain decomposition in the corresponding direction and expressing dependencies at the subdomain level.

Another overhead cost which needs to be investigated concerns the memory consumption. For computing each subdomain column, we use temporary buffers. The size of the temporary buffers for storing $U_n(:,j)$ and other intermediate results in the column-wise Godunov routine is of the same order of magnitude as the size of a subdomain column. In the original version, the global number of temporary buffers depends on the number of threads. With a simplistic task-based approach, the $U_n(:,j)$ buffers are allocated in the READ kernels. Unfortunately, if all the kernels of the Godunov routine are moved into separate tasks, then all of the READ tasks might be executed upfront by the scheduler (this is the worst case scenario in terms of memory footprint). In such a case, the memory consumption of the solver will dramatically increase as the global number of temporary buffers now depends on the number of READ tasks. Generating tasks for blocks of columns mitigates the problem but this problem can also be overcome by modifying the division of the algorithm between tasks as we shall see in the next paragraph.

Figures 6 and 7 present the algorithm dependencies more finely by distinguishing the operations involving interface cells. The dependencies between subdomains only concern cells at the interfaces. To compute the new grid values of a subdomain, interface cells of its two neighbours have to be read before those cells get modified by the neighbours. If interface cells are copied separately in a dedicated shared buffer by READ tasks, the other column-wise temporary buffers can be local to the COMPUTE tasks and the global memory overhead is limited to the size of the interfaces. One can notice that this task description of the algorithm is similar to the dependencies of a distributed memory domain decomposition : A READ kernel corresponds to an asynchronous communication of ghost cells between domains and tasking achieves the overlapping of communication with computations.

## 6. Implementation

The dependency graph of Figure 7 exhibits the full amount of available parallelism; however, since our initial goal was to facilitate the implementation of the "Coarse-Grain" version of HYDRO, we decided to implement a simplified version of the algorithm. We joined the tasks which are reading interface cells by twos as shown in Figures 8 and 9. A unique task is now responsible for reading all the interface cells between two domains. This modification reduces the potential parallelism but we believe that it is not consequential: The number of tasks can indeed be tuned by modifying the domain decomposition and furthermore, most of the time is spent in the COMPUTE kernel. This modification also slightly increases data locality access. After simplifying the READ kernels, every WRITE kernel directly depends on a single COMPUTE kernel and the dependency graph can be further simplified by merging COMPUTE and WRITE kernels.

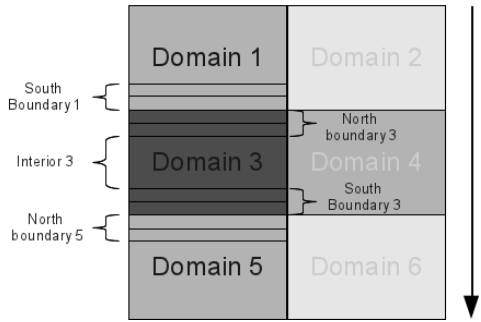
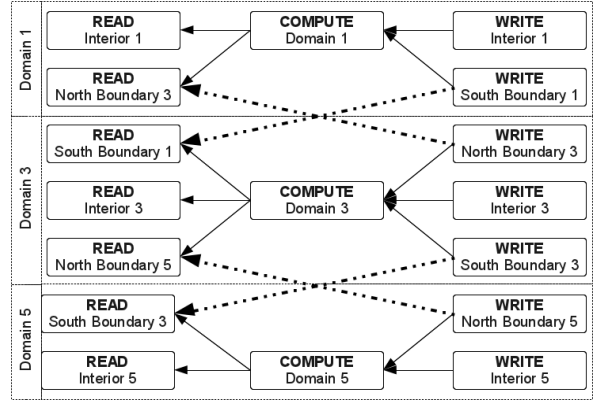

**Figure 6:** Description of the domain interfaces.

**Figure 7**: Fine description of the dependencies between computational kernels.

Figure 10 shows the algorithm in the column direction. In the subroutine *godunov*, we generate as many tasks as there are subdomains per row of the domain decomposition. Each task is responsible for the processing of a column of subdomains and there are no dependencies among these tasks. Inside of each of these tasks, two new set of subtasks is created: One set is for reading the interface cells (one task per interface between domains) and the other set is responsible for the computation and the update of the cell values (one task per subdomain). Dependencies among those tasks are expressed using the DEPEND clause of the OpenMP TASK construct.

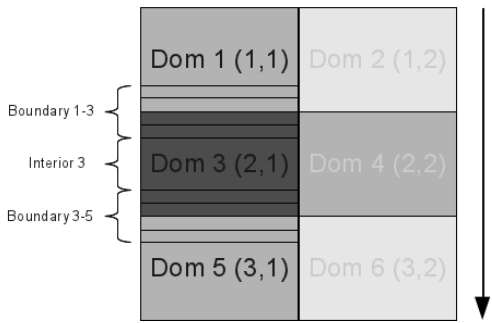
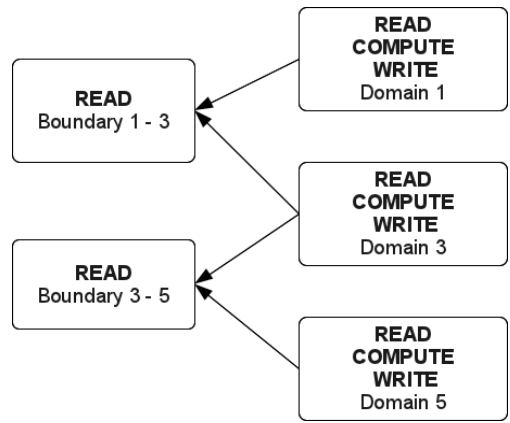

**Figure 8:** The domain decomposition with the domain coordinates and a simplified description of the interfaces.

**Figure 9**: Dependency graph of the implemented algorithm.

```
U      : 2D grid buffer for conservative variables.
coord2D : 2D array describing the domain decomposition,
          ie. coord2D(i,j) holds the information of the domain at coordinates (i,j)
          The coordinates of first domain is (1,1).

! Godunov update in the column direction
subroutine godunov(U, coord2D)
      !$ OMP DO
      do j=1,size(coord2D,2)
            !$ OMP TASK shared(U, coord2D)
                  ! processing one column of domains
                  call godunov_domaincol(j, U, coord2D)
            !$ OMP END TASK
      end do
      !$ OMP END DO
end subroutine
```

```
! processing one column of the domain decomposition
subroutine godunov_domaincol(j, U, coord2D)
        interface(i,:,:) : buffer storing the interface shared by domains (i,j) and (i+1,j).
        jmin = coord2(1,j)%jmin ! first column of domain (1,j)
        jmax = coord2(1,j)%jmax ! last column of domain (1,j)

        ! for each domain interface
        do i=1,size(coord2D,1)-1
                ! READ interfaces between domains
                !$ OMP TASK shared(U, coord2D) depend(out:interface(i,:,:))
                        ! index of the last line of domain (i,j)
                        imax = coord2(i,j)%imax
                        ! copy interface cells
                        interface(i,:,:) = U(imax-1:imax+2,jmin:jmax)
                !$ OMP END TASK
        end do

        ! for each domain of the domain decomposition column
        do i=1,size(coord2D,1)
                ! READ subdomain interior; COMPUTE and WRITE on the entire subdomain
                !$ OMP TASK shared(U, interface, coord2D) &
                !$ OMP depend(in:interface(i-1,:,:)) depend(in:interface(i,:,:))
                        ! process domain (i,j) (the godunov_domain routine is shown in Figure 5)
                        call godunov_domain(i, j, U, interface(i-1), interface(i+1), coord2D)
                !$ OMP END TASK
        end do
end subroutine
```

**Figure 10:** Godunov update in the column direction using OpenMP tasks.

### 7. Results

Figures 11, 12 and 13 show a scalability evaluation of the three OpenMP versions of HYDRO. The code is available on the HYDRO repository [1]. The experimental platform is the supercomputer Ada (CNRS/IDRIS-GENCI) composed of IBM System x3750 M4 compute nodes. This machine has a NUMA architecture and each node is a 4-socket system powered with 8-core Intel Sandy Bridge E5-4650 processors running at 2.7 GHz. We used the Intel Compiler and OpenMP runtime system. The test case is a point explosion problem using a 10000x10000 grid and 10 steps. For this experiment, we used the same number of domains in the "Coarse-Grain" and the "OpenMP tasks" version (one domain per thread).

The scalability of the "Fine-Grain" version of HYDRO suffers from a lack of data locality (one thread is responsible for the computation of different chunks of the global domain during the two successive 1D Godunov sweeps) but also from false-sharing among threads: In one of the direction, multiple threads update cells that are contiguous in memory resulting in cache misses when new cell values are written back in U (Step 3 of Figure 5).

The scalability of the "Coarse-Grain" version is nearly perfect, thanks to a very good use of the cache and the data locality. Threads work fully in parallel on their own part of the data and no time is wasted at the barriers. This performance comes at the cost of a greater code complexity.

The scalability of the new "OpenMP tasks" version is not as good as the finely tuned "Coarse-Grain" version since we are no longer in control of the affinity between threads and subdomains. But the results are very satisfactory, especially when compared to the "Fine-Grain" grain version that has the same problem of memory affinity.

The OpenMP specifications do not define how the scheduling of tasks should be done by runtime systems [10] and very little information is available concerning the actual strategy of our OpenMP library. Task scheduling and data placement over NUMA architectures is an active research field and several strategies have been proposed. For example, a hierarchical scheduling strategy with work-stealing is appropriate for recursive or irregular problems [11]. In HYDRO, the dependency direction changes between successive Godunov sweeps and it is difficult to generate a task hierarchy. In particular, tasks working on different subdomains have to be sibling tasks (i.e. tasks that are child tasks of the same task region) because there are dependencies among them. Therefore, keeping a good affinity between threads and subdomains on both of the Godunov sweeps seems difficult with a hierarchical scheduler. Another scheduling strategy consists in using the task data dependence information of OpenMP directives to guide the scheduling of tasks [12]. In HYDRO, such schedulers would be well suited because it is easy to identify which subdomain each task is working on. As for now, it is not possible to select the task scheduler strategy at the user level but it would be useful to be able to give hints to the scheduler in the same way we control the scheduling of parallel loop or the CPU affinity of threads in OpenMP.

|    | Time (s) | Speedup | Efficiency |
|----|----------|---------|------------|
| 1  | **389.21** | 1.00  | 100.00 %   |
| 2  | **202.49** | 1.92  | 96.11 %    |
| 4  | **118.40** | 3.29  | 82.18 %    |
| 8  | **97.42**  | 4.00  | 49.94 %    |
| 16 | **60.34**  | 6.45  | 40.31 %    |
| 32 | **29.72**  | 13.10 | 40.93 %    |

**Figure 11**: Scalability of the "Fine-Grain" version of HYDRO.

|    | Time (s) | Speedup | Efficiency |
|----|----------|---------|------------|
| 1  | **377.15** | 1.00  | 100.00 %   |
| 2  | **188.41** | 2.00  | 100.09 %   |
| 4  | **98.63**  | 3.82  | 95.60 %    |
| 8  | **52.18**  | 7.23  | 90.35 %    |
| 16 | **24.72**  | 15.26 | 95.36 %    |
| 32 | **11.91**  | 31.67 | 98.96 %    |

**Figure 12**: Scalability of the "Coarse-Grain" version of HYDRO.

|    | Time (s) | Speedup | Efficiency |
|----|----------|---------|------------|
| 1  | **376.81** | 1.00  | 100.00 %   |
| 2  | **188.99** | 1.99  | 99.69 %    |
| 4  | **98.65**  | 3.82  | 95.49 %    |
| 8  | **56.72**  | 6.64  | 83.04 %    |
| 16 | **32.96**  | 11.43 | 71.45 %    |
| 32 | **20.01**  | 18.83 | 58.85 %    |

**Figure 13**: Scalability of the "OpenMP tasks" version of HYDRO.

## 8. Summary and Conclusion

In this paper, we presented a first implementation of HYDRO using the new OpenMP tasking model. We implemented the synchronization algorithm of the "Coarse-Grain" version of HYDRO and compared the scalability of this new version with more traditional OpenMP implementations. The tasking model allows expressing a complex algorithm with ease and using tasks was less intrusive than implementing the algorithm with busy-waiting. We plan to extend this work with a more complete analysis of the performance and by investigating specialized task-based programming environments and runtime systems.

**Acknowledgements**

This work was financially supported by the PRACE project funded in part by the EUs 7th Framework Programme (FP7/2007-2013) under grant agreement no. PRACE-2IP: 283 493.